Electronic phase diagram of $NdFe_{1-x}Rh_xAsO$

David Bérardan, Lidong Zhao, Loreynne Pinsard-Gaudart, Nita Dragoe

Institut de Chimie Moléculaire et des Matériaux d'Orsay (UMR CNRS 8182), Bât 410, Univ. Paris-Sud 11, 91405 Orsay, France

We report on the electrical resistivity, thermoelectric power and electronic phase diagram of rhodium-doped NdFeAsO. Rhodium doping suppresses the structural phase transition and spin density wave observed in the undoped material, and superconductivity emerges at x close to 0.05, despite the distortion of $FeAs_4$ tetrahedra induced by the large size difference between Rh and Fe elements. The $T_c(x)$ curve is dome-like, and the highest $T_c$ is reached at x = 0.1, with $T_c^{onset}$ = 18K. An upturn of the electrical resistivity above $T_c$ has been observed, with a Kondo like behaviour above $T_c$ and a Fermi-liquid behaviour close to room temperature.

**Introduction**

Since the discovery of superconductivity at 26K in $LaFeAsO_{1-x}F_x$ ("1111" compounds) in 2008 [1], an intense research activity has emerged dealing with the study of the oxypnictides, a new family of high-$T_c$ superconductors outside the cuprates family. In a matter of months after the first report of Kamihara *et al.*, the critical temperature of these compounds has been raised to over 50K by replacing lanthanum by smaller rare-earth elements [2]. For a recent review about the superconductivity in this materials family, see ref 3.

The parent compound LaFeAsO, which consists in alternating $La_2O_2$ and $Fe_2As_2$ layers, is not superconducting, but it exhibits a metallic behaviour with a structural transition from tetragonal to orthorhombic as well as a spin density wave, both emerging around 150K [4]. Upon doping, the structural transition and the spin density wave are both destroyed, and a superconducting state is observed [4]. This superconducting state can be induced by electron doping, with fluorine doping [1] or oxygen vacancies formation [5] on the oxygen site, or with thorium doping on the rare-earth site [2]. It can also emerge from a hole doping with strontium on the rare-earth site [6]. More surprisingly, it has been shown that in addition to the electron or hole doping in the $La_2O_2$ layers, a partial substitution of iron by cobalt [7] or nickel [8] in the $Fe_2As_2$ layers also leads to the emergence of superconductivity, which is very different of the cuprates behaviour where the superconducting state is very sensitive to the presence of impurities on the copper site. Indeed, the electronic phase diagram proposed by Wang *et al.* for $LaFe_{1-x}Co_xAsO$ exhibits a dome-like curve for $T_c$ as a function of Co content [9], which resembles that of $LaFeAsO_{1-x}F_x$ [1]. However, it seems that hole doping cannot be induced by substitutions in the $Fe_2As_2$ layers [10, 11]. Recently, Lee *et al.* have reported the effect of Ru doping in optimally doped $NdFeAsO_{0.89}F_{0.11}$ [12]. These authors showed that the rate of $T_c$ decrease induced by Ru-doping is very small, despite the structural distortion induced by the doping. Moreover, we have shown that despite the large size difference between Fe and Rh elements, superconductivity can be induced by rhodium doping in NdFeAsO [13]. In this paper, we report on the transport properties and the electronic phase diagram of $NdFe_{1-x}Rh_xAsO$.

## Experimental

Samples with nominal composition $NdFe_{1-x}Rh_xAsO$ ($0.025 < x < 0.20$) were prepared by a solid state reaction route, using Nd, As, Fe and Rh metals and $Fe_2O_3$ powder. All handlings were made in an Ar-filled glovebox with less than 1 ppm $O_2$ and $H_2O$. NdAs alloy was first obtained by heating Nd and As under pure Ar in a closed silica tube at 900°C. The single phase nature of the alloy was confirmed by X-ray diffraction. NdAs was then carefully mixed in stoichiometric ratio with Fe, $Fe_2O_3$ and Rh, and the resulting powder was pressed into 2x3x12 $mm^3$ bars under 200 MPa. These bars were heated two times with intermediate grinding and pressing, at 1150°C during 48h under argon in closed silica tubes. X-ray diffraction characterization was performed using a Panalytical X'Pert diffractometer with a Ge(111) incident monochromator and a X'celerator detector. The XRD patterns were analyzed using the Rietveld method with the help of the FullProf software [14]. The thermoelectric power was measured by a differential method with two T-type thermocouples, by using the slope of the $\Delta V$–$\Delta T$ curve with thermal gradients along the samples up to about 0.3 K/mm, in a laboratory made system. The electrical resistivity was obtained between 2K and 300K and between 0 and 9 T by a DC four wires method using a Quantum Design Physical Properties Measurement System (PPMS) and silver paste for the electrical contacts. All transport measurements were performed in a direction perpendicular to the pressing direction. Magnetic susceptibility was measured down to 5K using a quantum-design MPMS under 20 Oe.

## Results and discussion

Figure 1 shows the XRD patterns of $NdFe_{1-x}Rh_xAsO$ compounds (x = 0.025 to 0.2). All major Bragg peaks can be indexed using a tetragonal unit cell with structure type ZrCuSiAs [15], represented in the inset of figure 2, indicating that the samples are almost single phase. An example of Rietveld refinement is shown in figure 2, which also shows the positions of the Bragg reflexions of the 1111 phase. The typical R values for all the refinements are $R_F$ ~ 3%, $R_{Bragg}$ ~ 3% and $R_{wp}$ ~ 12%. Less than 3% secondary phase is observed, which corresponds to $Nd(OH)_3$, that results from $Nd_2O_3$ minority phase under air. A partial preferential orientation of the grains along the [0 0 1] direction has been observed, which is related to the microstructure of these compounds and most probably originates from the uniaxial pressing of the powder.

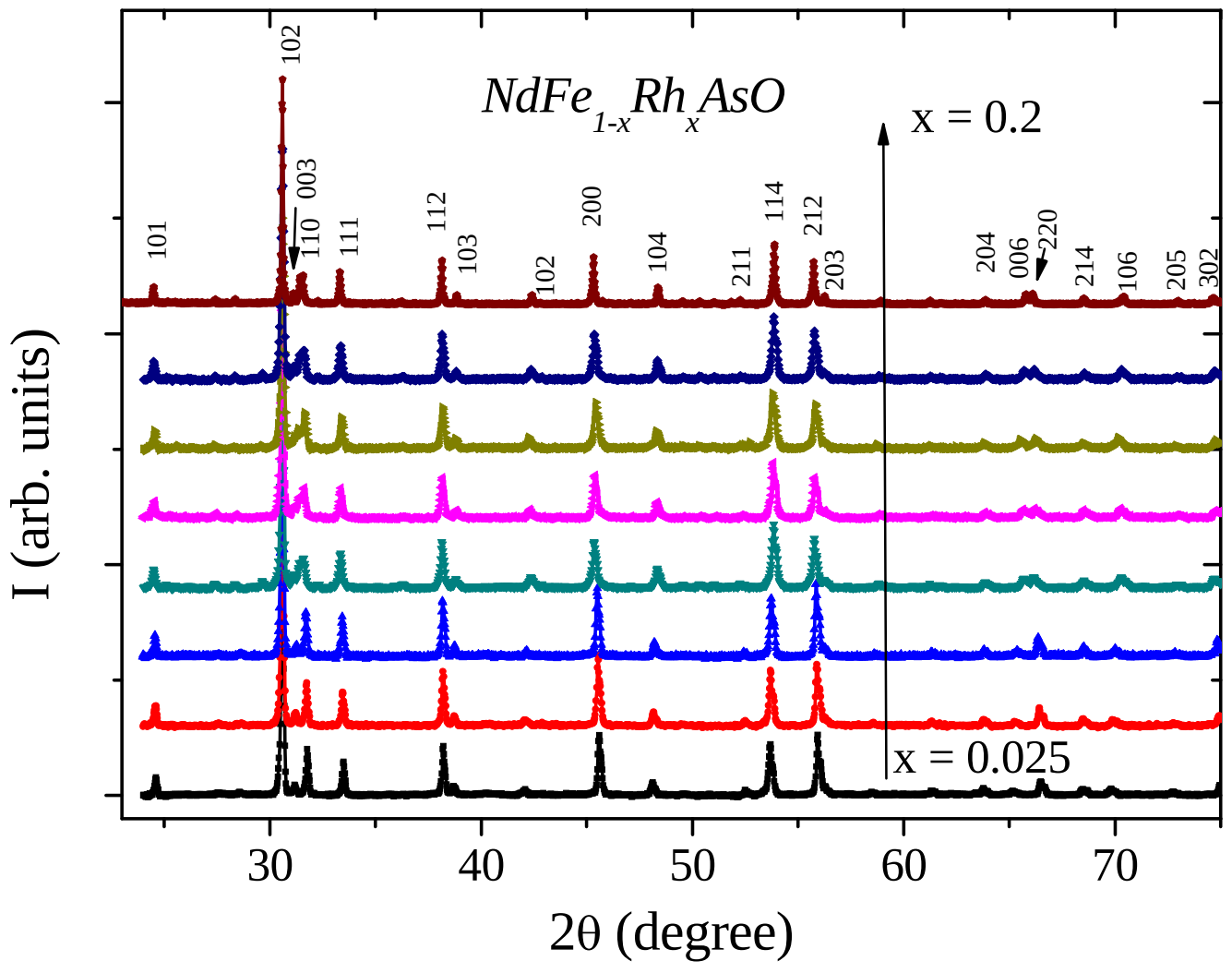

Figure 1: XRD patterns of $NdFe_{1-x}Rh_xAsO$ compounds, with x = 0.025 (bottom) to x = 0.2 (top)

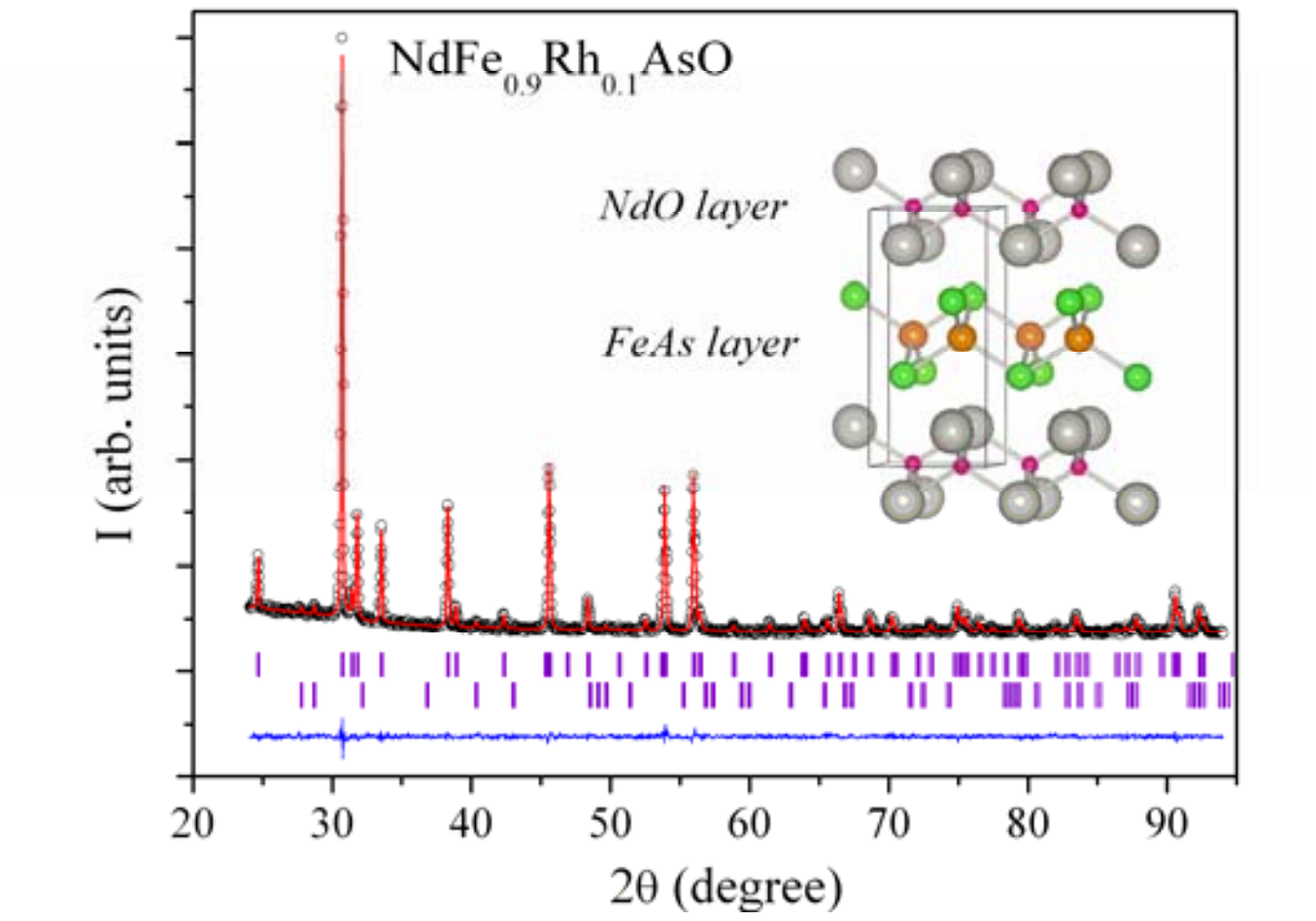

Figure 2: XRD pattern and Rietveld refinement of $NdFe_{0.9}Rh_{0.1}AsO$. Minor impurity peaks correspond to $Nd(OH)_3$. The crystal structure is also plotted showing the FeAs and NdO layers (inset).

The structural parameters determined using the Rietveld refinements are summarized in figure 3 as a function of the rhodium fraction. A significant increase of *a* is observed with increasing rhodium fraction, which is consistent with the atomic radius of Rh being larger than that of Fe [16]. This linear trend confirms that the substitution of iron by rhodium in the NdFeAsO unit cell occurs. Despite the increase of the lattice parameter *a* with increasing Rh fraction, the unit cell volume is almost unaffected by the substitution, due to the simultaneous decrease of *c* that results in a reduced interlayer distance. Schematically, the NdFeAsO unit cell can be pictured as alternating $[Nd_2O_2]^{2+}$ and $[Fe_2As_2]^{2-}$ layers (see figure 2). The substitution of iron by rhodium leads to an increase of the electrons density in the metal-pnictide layer following $[Fe_{2-2x}Rh_xAs_2]^{(2+2x)-}$, which increases the interlayer Coulomb attraction and decreases the interlayer distance and therefore the lattice parameter *c*. A similar behaviour has already been reported in Co-doped LaFeAsO, where the substitution of iron by cobalt leads to a strong decrease of the *c* parameter whereas *a* remains almost unchanged [9,7]. This increase of the carriers concentration with Rh-doping is confirmed by our thermoelectric power measurements, see later. Due to the simultaneous decrease of *c* and increase of *a*, the distortion of the $FeAs_4$ tetrahedra increases. As can be seen in figure 3, the difference between the two characteristic AsFeAs angles of the $FeAs_4$ tetrahedra increases gradually with the rhodium substitution. On the contrary, fluorine doping leads to a decrease of the distortion of the tetrahedral as compared to NdFeAsO.

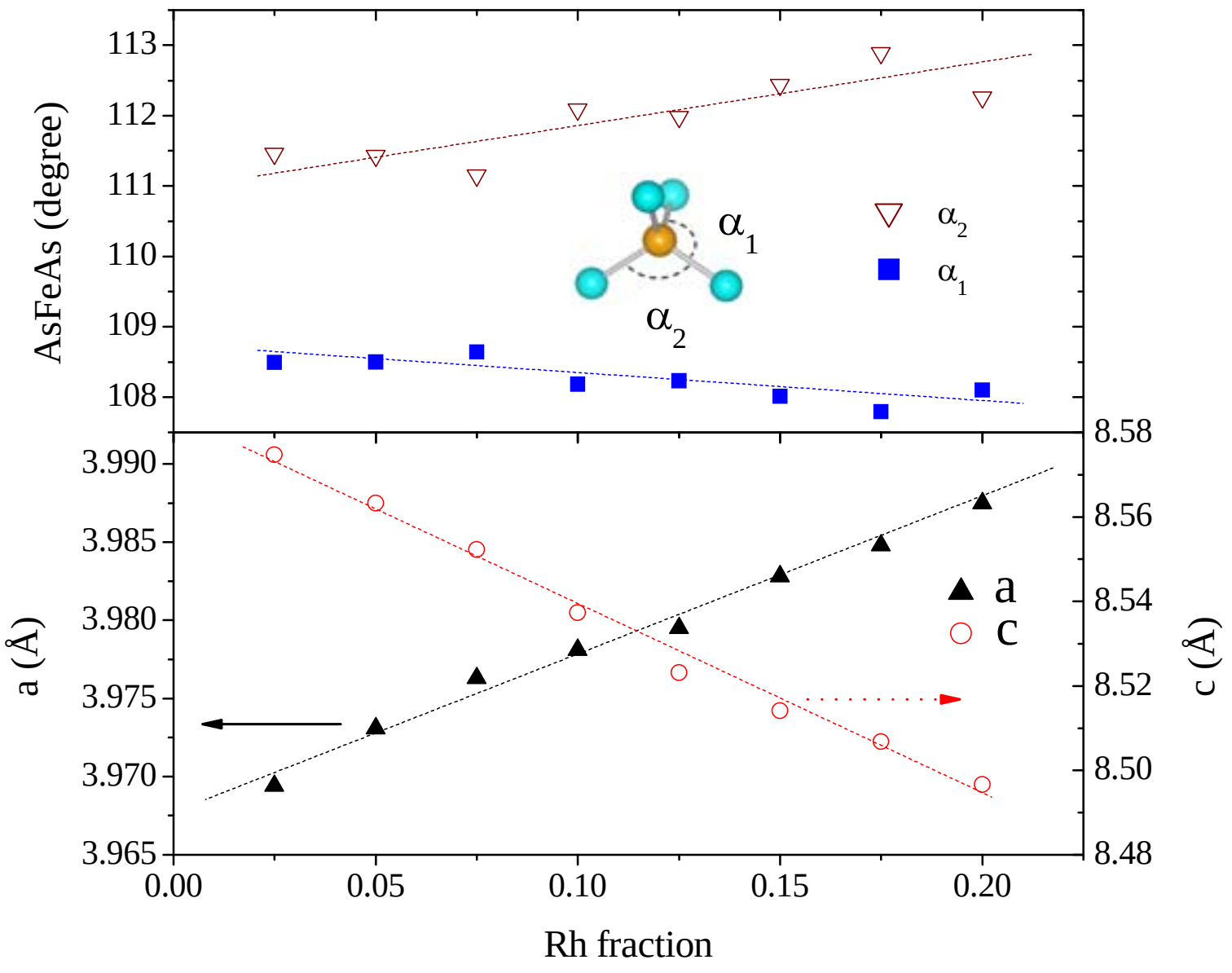

Figure 3: Evolution with the rhodium fraction in $NdFe_{1-x}Rh_xAsO$ of the lattice parameters (bottom), and of the angles of the $FeAs_4$ tetrahedra (top). Lines are visual guides.

Figure 4 shows the temperature dependence of the electrical resistivity in the series $NdFe_{1-x}Rh_xAsO$ normalized to the room temperature values. The undoped parent compound is known to exhibit a strong anomaly of the electrical resistivity around 150K followed by a large decrease of the resistivity at lower temperature (see for example ref 17). Upon doping with 2.5 atom% rhodium, the anomaly shifts to a lower temperature ($T_{anom}$ ~ 95K) and becomes less pronounced. The origin of this anomaly is probably the same as is the undoped compound, where it has been attributed to a spin density wave (SDW) and a structural transition from tetragonal at room temperature to orthorhombic at low temperature. With a further increase of the rhodium fraction ($x > 0.05$), no anomaly can be observed any longer, and an unambiguous superconducting transition emerges at low temperature. The inset of figure 3 shows the volumic magnetic susceptibility of $NdFe_{0.9}Rh_{0.1}AsO$, which exhibits the highest $T_c$ in the series, obtained in a zero-field cooled mode with an excitation field of 20 Oe, assuming a sample density equal to the theoretical one. A superconducting transition can be observed around 15K. From the diamagnetic signal at 5K, the shielding fraction can be estimated to about 20% of the volume of the sample, which confirms the bulk nature of the superconductivity.

The highest critical temperature observed in the $NdFe_{1-x}Rh_xAsO$ series, $T_c^{onset} \sim 18K$, is much lower than the one observed in $NdFeAsO_{0.88}F_{0.12}$ [18]. A similar difference has already been reported between fluorine doped and cobalt doped LnFeAsO [1,9]. Lee *et al.* have suggested that the highest $T_c$ is obtained when the $Fe_2As_2$ lattice forms regular tetrahedra [19]. Our results are consistent with this picture, with $FeAs_4$ tetrahedra being more distorted and $T_c$ being lower with Rh-doping than with F-doping. However, the maximum critical temperature observed in $NdFe_{0.9}Rh_{0.1}AsO$ is much lower that the one that would be expected using the $T_c$ vs. $\alpha_2$ dependence suggested in ref 19, which should be closer to 30K.

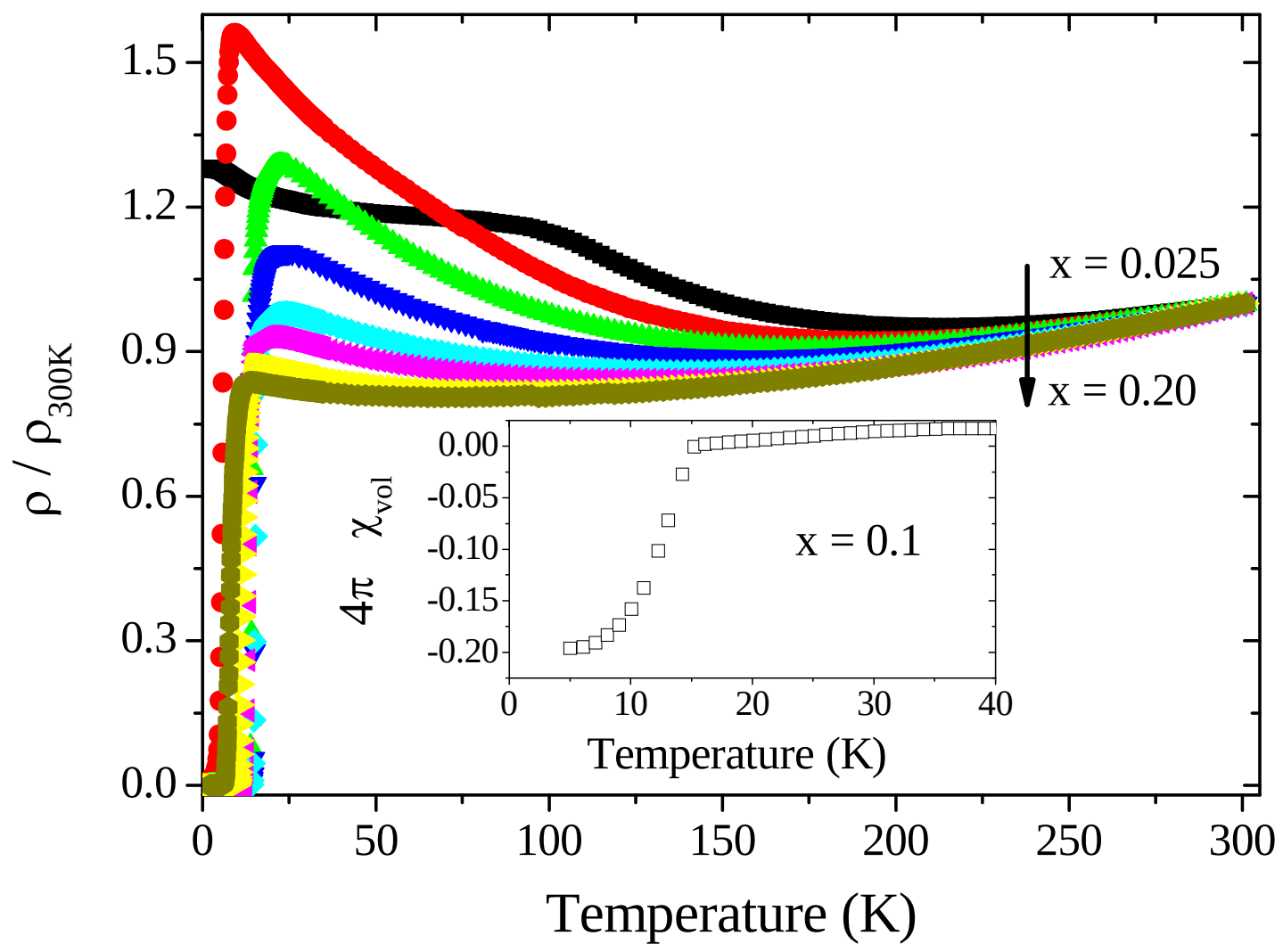

Figure 4: Temperature dependence of the normalized electrical resistivity in the series $NdFe_{1-x}Rh_xAsO$. Inset: Temperature dependence of the magnetic susceptibility of $NdFe_{0.9}Rh_{0.1}AsO$ recorded in a zero-field cooled mode at 20 Oe.

Two significant differences between the temperature dependence of the electrical resistivity of the fluorine doped sample [18] and of the rhodium doped ones (figure 4) can be underlined.

First, although both fluorine doped and rhodium doped samples exhibit a metallic electrical resistivity close to room temperature, the temperature dependence of the resistivity is different. $NdFeAsO_{0.88}F_{0.12}$ behaves like a "strange metal", with the exponent $x$ of $\rho \sim T^x$ that changes from $x>1$ to $x<1$ when the temperature exceeds about 150K [20, 21]. On the other hand, rhodium doped samples exhibit a Fermi-liquid behaviour with $\rho \sim \rho_0 + AT^2$, which evidences enhanced electron-electron interactions. Figure 5 shows the temperature dependence of the normalized electrical resistivity in the series $NdFe_{1-x}Rh_xAsO$, with the temperature scale in $T^2$. A linear trend can be observed for every sample above 200-220K (0.05<x<0.2). These observations, which are in good agreement with the results reported in the series $LaFe_{1-x}Ni_xAsO$ [22], are not fully consistent with the electronic phase diagram suggested by Hess *et al.* [23] for FeAs superconductors, who indicated that the Fermi-liquid behaviour should be a characteristic feature of the overdoped regime.

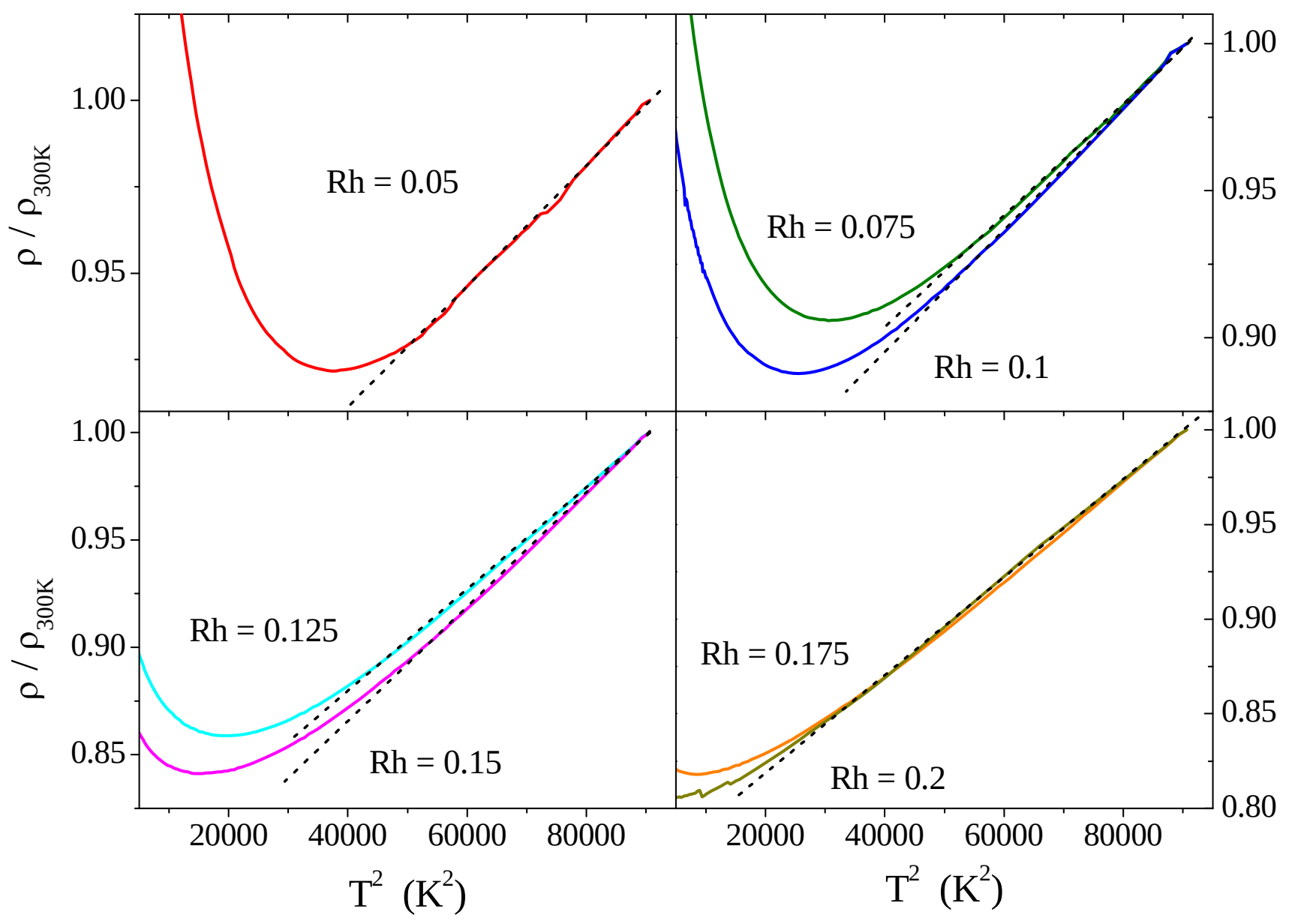

Figure 5: Temperature dependence of the normalized electrical resistivity in the series $NdFe_{1-x}Rh_xAsO$, temperature scale in $T^2$. The black dashed lines are linear fits.

Secondly, for the fluorine-doped sample, the resistivity in the normal state is typical of a metal from $T_c$ to room temperature. On the contrary in the rhodium doped compounds, whereas the resistivity is also metallic around room temperature, a minimum can be observed at $T_{min}$ followed by an upturn of the resistivity that exhibits a semiconductor-like behaviour between $T_{min}$ and $T_c$ ($0.05 < x < 0.2$). The temperature of the minimum of the resistivity $T_{min}$ shifts to lower temperatures when the rhodium fraction is increased. Moreover, the maximum of resistivity above $T_c$ becomes less pronounced for higher Rh-fractions. A similar behaviour has already been reported in Ni-doped [8] and Co-doped LnFeAsO [7,9]. A much less pronounced upturn of the resistivity has also been reported in some F-doped LaFeAsO samples with small fluorine content [1] whereas it does not seem to be present in optimally doped samples or other rare-earth than lanthanum (ref 18, 22, 23, 24). At the first glance, it seems that this metallic to semiconductor-like transition could be simply explained by the local structural disorder induced in the conducting $Fe_2As_2$ layer by rhodium doping. With increasing rhodium fraction, the increased carriers concentration would lead to a more metallic behaviour that would eventually hide the upturn. However, this simple picture is not consistent with the results recently published by Lee *et al* [12]. These authors reported the effects of ruthenium doping on the electrical transport behaviour of fluorine doped $NdFeAsO_{0.89}F_{0.11}$. Whatever the Ru fraction, no maximum of the resistivity has been observed and all $NdFe_{1-y}Ru_yAsO_{0.89}F_{0.11}$ samples are metallic above $T_c$, which seems to rule out the possible role of the structural disorder. Moreover, this upturn cannot be linked to thermal excitation of carriers through a band-gap, as band structure calculations and photoemission spectroscopy experiments have unambiguously proved the metallic nature of this family of compounds (see for example ref 24).

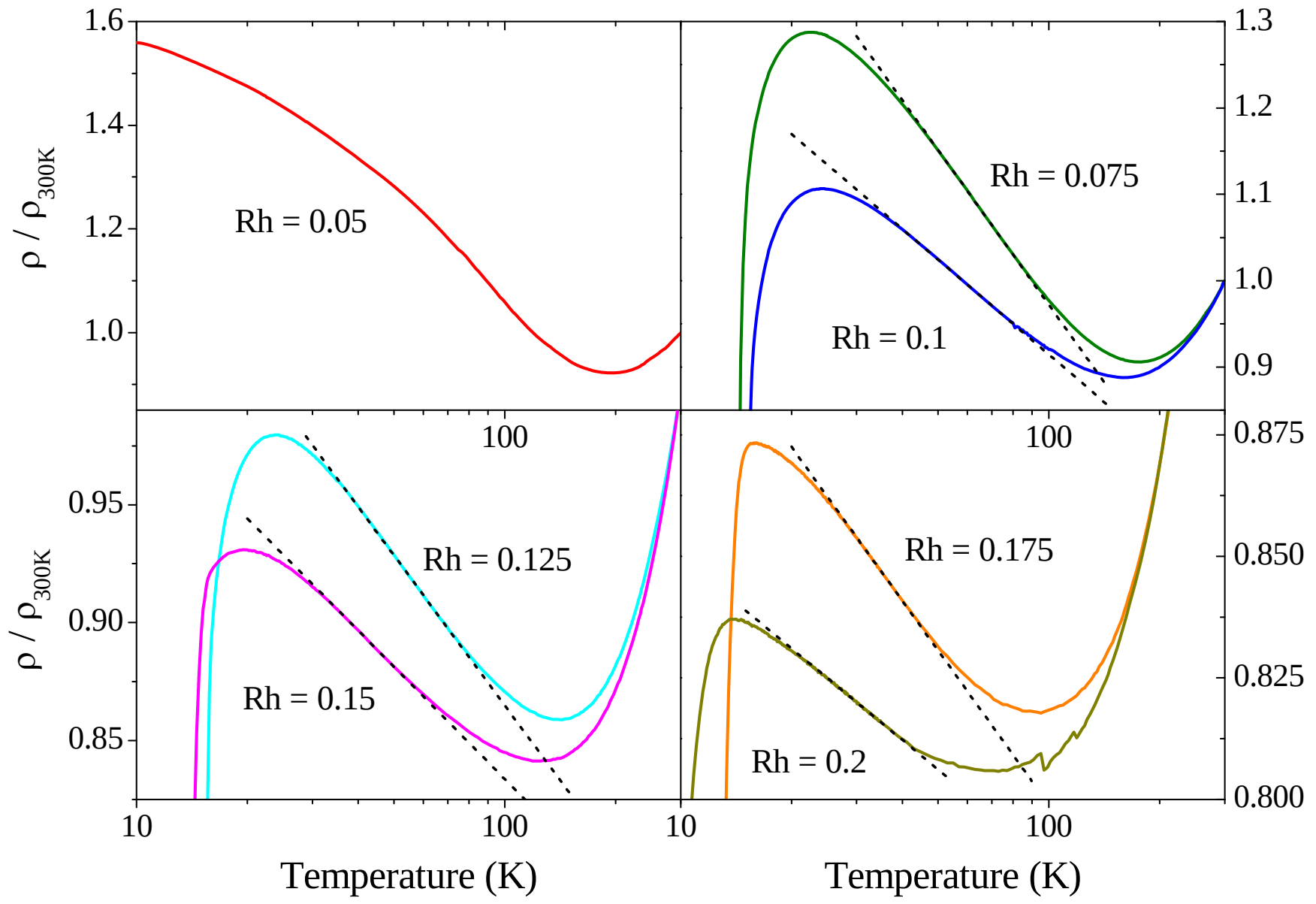

Figure 6: Temperature dependence of the normalized electrical resistivity in the series $NdFe_{1-x}Rh_xAsO$, temperature scale in log(T). The black dashed lines are linear fits.

Recently, Cao *et al*. have suggested that the resistivity upturn above $T_c$ could originate from Kondo effect [22,25]. In their picture, Ni doping in the $Fe_2As_2$ layer of LaFeAsO not only induces itinerant carriers, but also stabilized localized moment, in agreement with band structure calculations [26]. Indeed, the coexistence of itinerant charge carriers and local moments and a resistivity upturn following a log(T) behaviour is often characteristic of Kondo effect [27]. A similar log(T) dependence of the normal state resistivity as also be observed by Tropeano *et al.* in $Fe_{1+x}Te_{1-x}Se_x$ [28], coupled to a $B^2$ dependence of the magnetoresistivity. Moreover, the temperature dependence of the reported magnetoresistivity is well described using a Kondo formalism. In this latter case, the magnetic impurities originate from the excess of Fe, which provides localized magnetic moments [29, 30].

Figure 6 shows the temperature dependence of the normalized electrical resistivity in the $NdFe_{1-x}Rh_xAsO$ series, with the temperature scale in log(T). Although no linear trend can be observed with x=0.05 in any temperature range, all samples with x > 0.075 exhibit a linear behaviour in a narrow temperature range (black dashed lines). The absence of linear behaviour with x = 0.05 could be connected to a reminiscence of the spin density wave, as this compound lies close to the boundary of the superconducting area of the phase diagram. The width of this log(T) area decreases with increasing Rh fraction, from about 65K with x = 0.075 to about 25K with x = 0.2. This decrease is connected to the electron doping induced by the substitution of Fe by Rh. Indeed, band structure calculations have shown that the density of state at the Fermi level $N(E_F)$ is reduced upon doping in the LnFeAsO system [26]. As the Kondo temperature decreases when $N(E_F)$ decreases, increasing Rh fraction leads to a decrease of $T_{min}$. The upturn of the electrical resistivity above $T_c$ is more pronounced in Rh-doped samples than in the F-doped ones (where it is not always present), because the substitution of iron by rhodium increases not only the carrier concentration but could also induce the presence of local moments in the $Fe_2As_2$ layer, similarly to Ni doping in LaFeAsO or Fe excess in $FeTe_{1-x}Se_x$, which in turn increases the electrons scattering rate. This observation is also consistent with the highest observed $T_c^{onset}$ being lower than the one expected considering the distortion of the $FeAs_4$ tetrahedra, as the local moments induced by Rh doping could also act as pair-breaker and lower the critical temperature. Nevertheless, the resistivity upturn above Tc could also originate from weak-localization effects linked to the 2D character of this material, and no definitive conclusion can be drawn at the moment.

An electronic phase diagram of the series $NdFe_{1-x}Rh_xAsO$ based on the electrical resistivity data is plotted in figure 7. The SDW area is very narrow. The frontier is not well defined, and from our data, it is not clear if there is coexistence of the SDW and of superconductivity in $NdFe_{0.95}Rh_{0.05}AsO$. For Rh fractions higher than 0.05, the SDW is clearly suppressed, and a superconducting state emerges at low temperature, with a dome-like $T_c$(Rh) curve, characteristic of the FeAs superconductors. Although the details are different, this dome-like evolution is similar to the one reported for Co-doped LnFeAsO [7,9]. However, contrary to the Co-doped systems, superconductivity is not destroyed up to x > 0.2. For x > 0.075, the normal state resistivity exhibits a Fermi-liquid behaviour close to room temperature, that changes to a “Kondo-like” temperature dependence above $T_c$, with $T_{min}$ that decreases with increasing rhodium fraction.

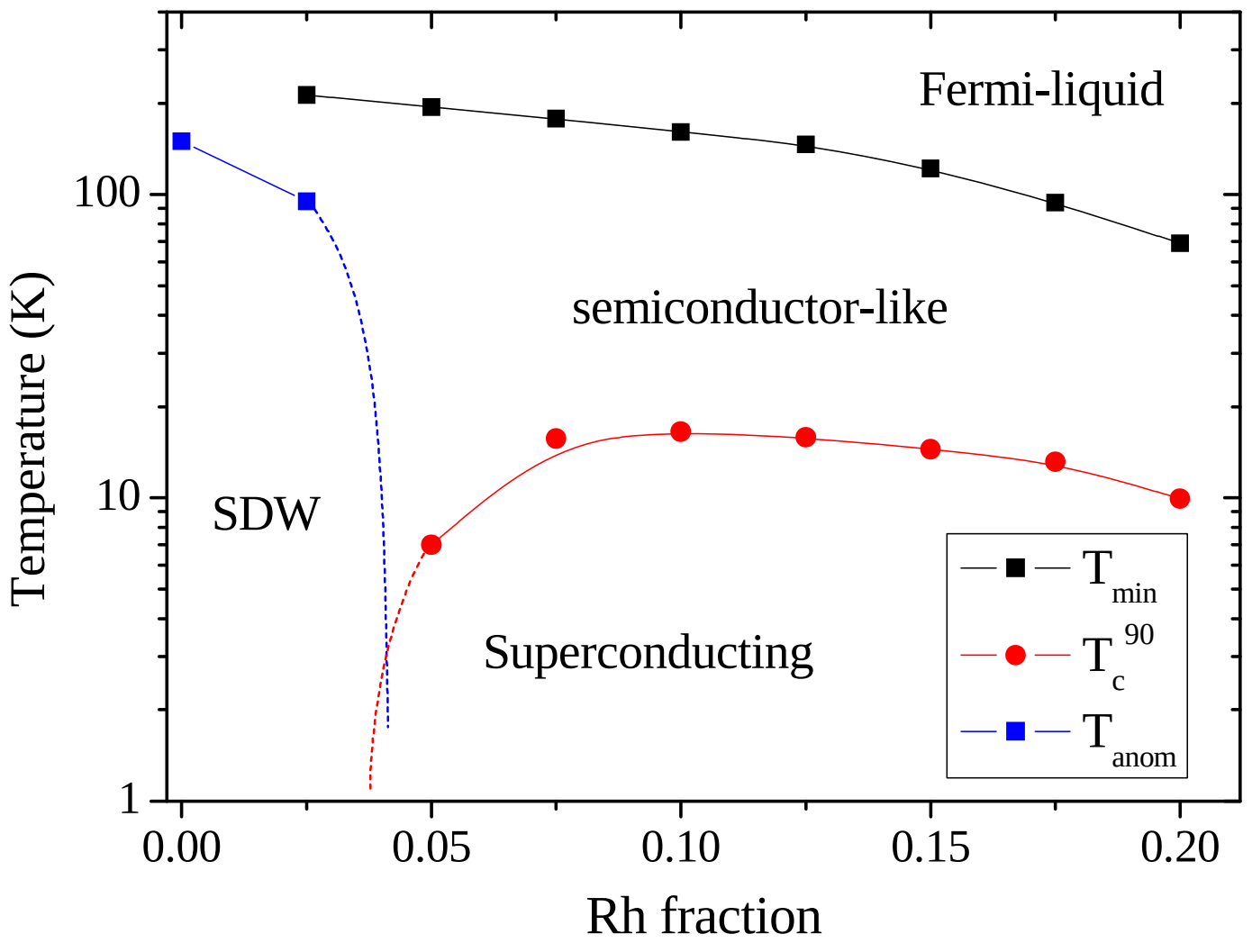

Figure 7: Electronic phase diagram for $NdFe_{1-x}Rh_xAsO$ (vertical axis is in logarithmic scale). The point at x = 0 is taken from ref 19.

We have compared the temperature dependence of the electrical resistivity under several magnetic fields up to 9 T for $NdFe_{0.9}Rh_{0.1}AsO$, which exhibits the highest $T_c$ value in the $NdFe_{1-x}Rh_xAsO$ series, in order to estimate the upper critical field $H_{c2}$. All curves are plotted in figure 8. $T_c^0$ corresponds to the zero resistivity temperature, defined as the temperature where the resistivity goes below the sensibility limit of the PPMS. As can be seen in figure 8, $T_c^{90}$ decreases very weakly with the magnetic field, whereas $T_c^0$ decreases more rapidly leading to a widening of the superconducting transition, characteristic of type II superconductivity (the field dependence of $T_c^0$ and $T_c^{90}$ are plotted in the inset of figure 8). It is noteworthy that $T_c^0$ is still higher than 4K even under a magnetic field as high as 9 T, which evidences the very robust character of the superconducting state in this family of materials. Without magnetic field, the width of the superconducting transition is small. If we define $\Delta T_c$ as $T_c^{90} - T_c^{10}$, $\Delta T_c$ is a low as 1.8K, which is much smaller than the value reported for cobalt doped LaFeAsO [7] or that the one of fluorine doped NdFeAsO [31].

The slope $(dH_{c2}/\, dT)_{|T=Tc}$ for $T_c^{90}$ is about -9 $T.K^{-1}$. Using the Werthamer-Helfand-Hohenberg formula $H_{c2}(0) = -0.693\ T_c\ (dH_{c2}/\, dT)_{|T=Tc}$ leads to $H_{c2}(0)$ of the order of 100 T, which is much higher than the values reported for Co-doped LnFeAsO [7,32], and of the same order as the value of $NdFeAsO_{0.88}F_{0.12}$ [31].

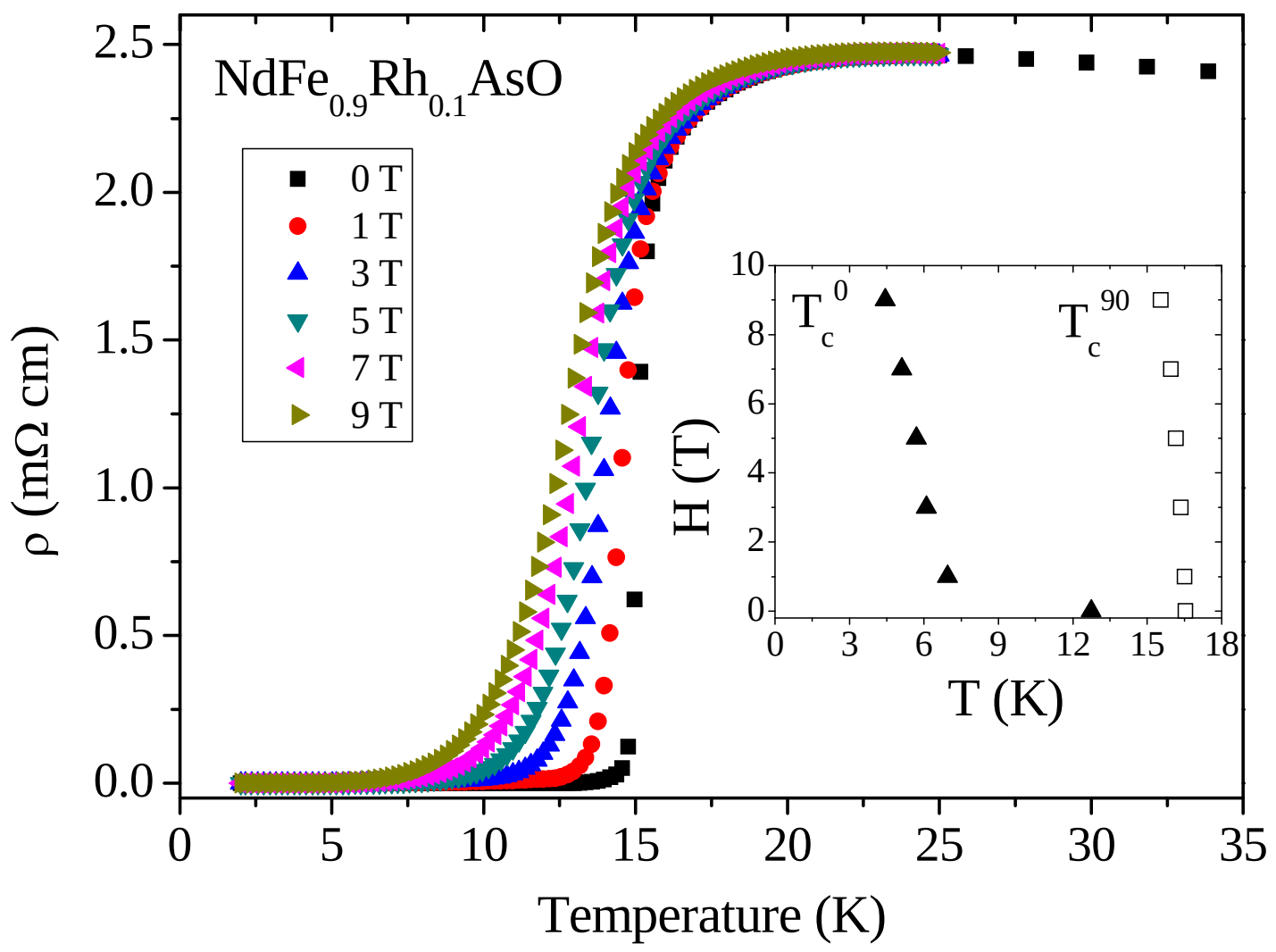

Figure 8: Field dependence of the superconducting transition in $NdFe_{0.9}Rh_{0.1}AsO$. Inset: Field dependence of $T_c^{90}$ and $T_c^{0}$.

In the superconducting 1111 compounds, a peak of the thermoelectric power above $T_c$ has been widely reported (see for example ref 33 or 34). Wang *et al.* have suggested that this enhancement of the thermoelectric power in superconducting compounds could be somehow correlated to the onset of superconductivity [9]. Figure 9 shows temperature dependence of the thermoelectric power in the series $NdFe_{1-x}Rh_xAsO$. All samples are n-type, which is consistent with an increase of the electrons concentration induced by the substitution of Fe by Rh. Band structure calculations have shown that 1111 compounds are multiband materials, and that several electron and hole pockets contribute to the electrical transport [35]. In a multiband system, the thermoelectric power is a mixture of the individual thermoelectric power of each contributing band following:

$$S_{tot} = \frac{\sum \sigma_i S_i}{\sum \sigma_i}$$

with $\sigma_i$ and $S_i$ being respectively the conductivity of the $i^{th}$ band and its contribution to the thermoelectric power (negative for an electron band, positive for a hole band). The thermoelectric power of undoped NdFeAsO has been reported by McGuire *et al.* [17]. Below about 150 K, the thermoelectric power is dominated by electron pockets and is positive. Then a sharp transition occurs at about 150 K, with denotes a strong evolution of the conduction mechanism with the structural transition and the emergence of the spin density wave. Above 150 K, the thermoelectric power is negative, up to room temperature. When iron is substituted by rhodium, the hole pockets are gradually suppressed, and the thermoelectric power becomes n-type in the whole temperature range. Although the sign of the thermoelectric power does not change, a reminiscence of the transition between electron-dominated and hole-dominated thermoelectric power can still be observed for x = 0.025. When the rhodium fraction is increased further, no feature (transition, bump or kink) can be observed and the temperature dependence of the thermoelectric power gradually approaches that of a metal. This evolution can also been explained considering a gradual suppression of the hole pockets and therefore a decrease of the “positive” contributions to the total thermoelectric power.

If we compare the thermoelectric power of F-doped samples [33] and that of Rh-doped samples, it is noteworthy that no significant peak can be observed above $T_c$ in superconducting $NdFe_{1-x}Rh_xAsO$ samples (see for instance x = 0.1 and x = 0.125, which correspond to the highest observed $T_c$). This behaviour is also slightly different from the one observed for Co-doped LnFeAsO [7,9]. Therefore, it seems that the electronic band structure is influenced by Rh-doping as compared to F- or Co-doping.

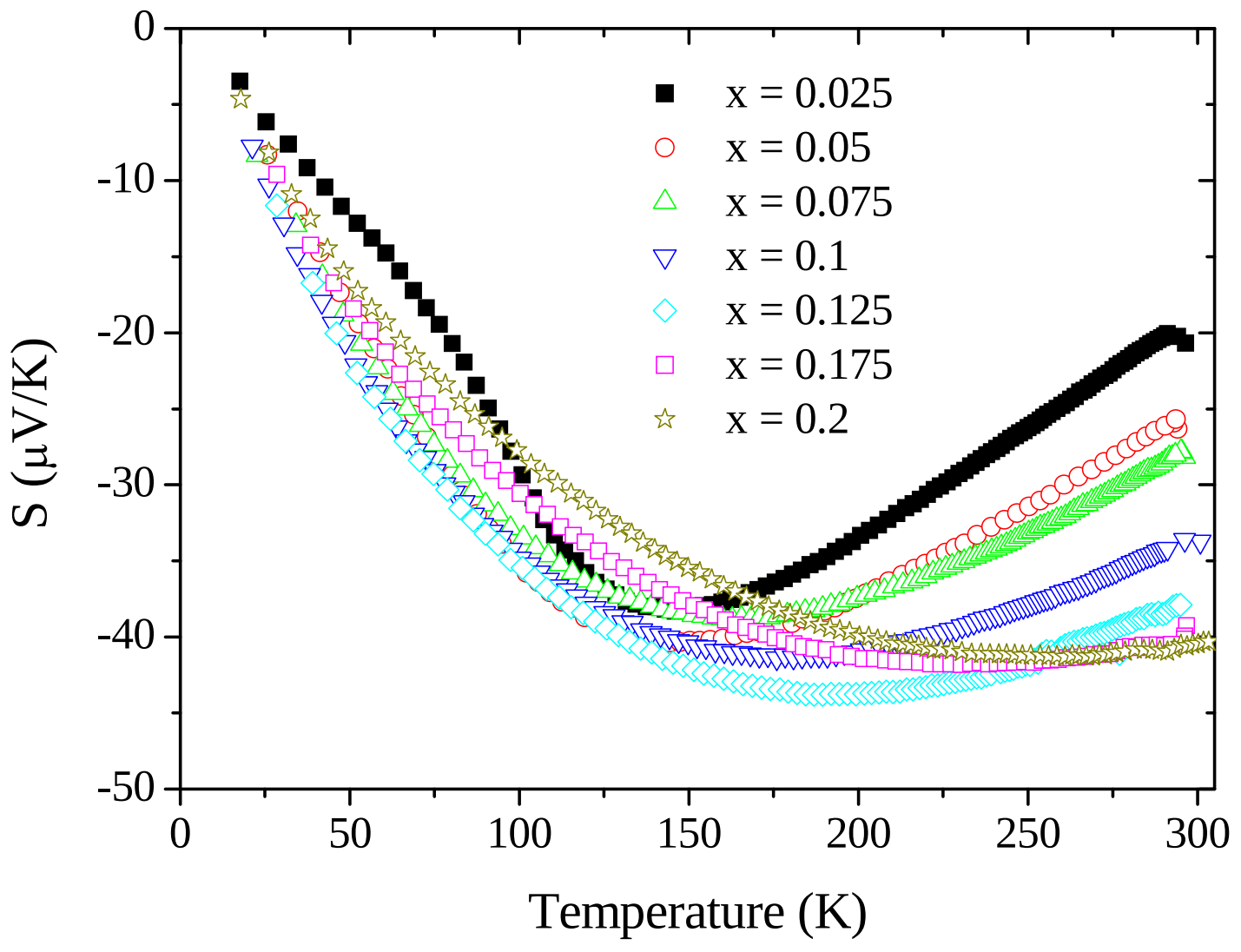

Figure 9: Temperature dependence of the thermoelectric power in the series $NdFe_{1-x}Rh_xAsO$.

In order to compare the behaviour of $T_c$ and S of Rh-doped samples to the ones of Co-doped samples as reported by Wang *et al.* [9], the evolution with the rhodium fraction of $T_c^{90}$, the room temperature thermoelectric power and the room temperature resistivity have been plotted in figure 10. No obvious link can be seen between the critical temperature and the thermoelectric power, and no anomalous enhancement of the thermoelectric power in the superconducting window can be seen. Therefore, this enhancement does not seem to be an universal feature of iron-based arsenic superconductors. However, the evolutions of the room temperature values of the resistivity and of the thermoelectric power can both be well explained taking into account an increase of the electrons concentration in the system due to rhodium doping. This raises into question the origin of the abnormal contribution to the thermoelectric power in F-doped iron-pnictides.

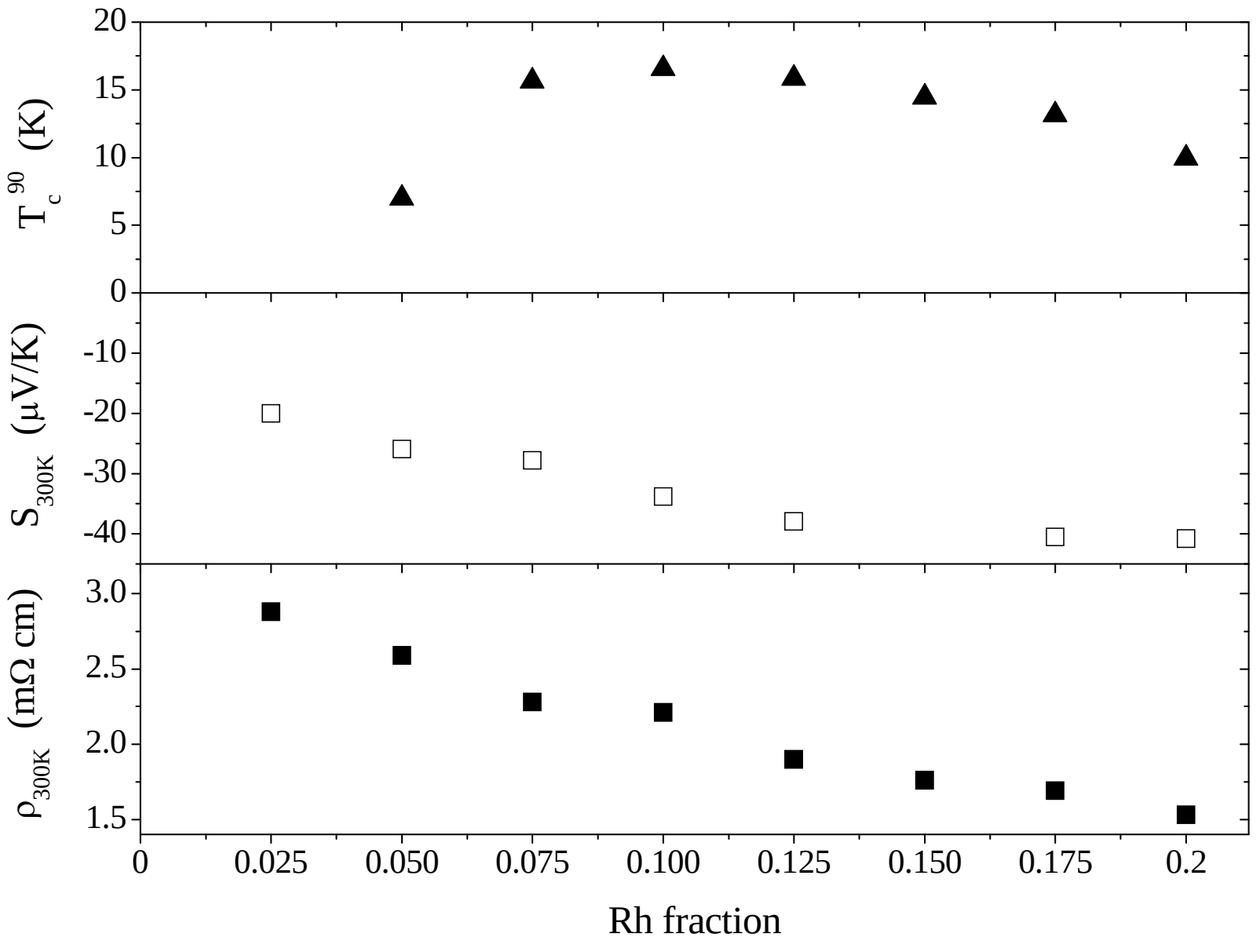

Figure 10: Evolution of the room temperature electrical resistivity (bottom), thermoelectric power (middle) and critical temperature $T_c^{90}$ with the rhodium fraction in $NdFe_{1-x}Rh_xAsO$.

## Conclusion

To sum up, our systematic study of the electrical transport properties in the series $NdFe_{1-x}Rh_xAsO$ have lead to establish the electronic phase diagram of Rh-doped NdFeAsO. The evolution with rhodium doping of the thermoelectric power as well as the room temperature value of the electrical resistivity indicates that rhodium doping on the iron site first leads to a suppression of the hole pockets at the Fermi level and then to an increase of the electrons concentration. The spin density wave and the structural transition are suppressed by Rh-doping and a superconducting state emerges for a rhodium fraction close to 0.05. Rh-doped NdFeAsO compounds exhibit a dome-like dependence of $T_c$, similar to F-doped compounds, although the highest value of $T_c$ is much lower with Rh-doping. This difference can be well explained taking into account both the distortion of the $FeAs_4$ tetrahedra and the pair-breaking effect of localized moments in the $Fe_2As_2$ layer induced by Rh-doping. These localized moments also possibly induce a scattering of the electrons above $T_c$ leading to a Kondo-like behaviour of the electrical resistivity, which is not observed in F-doped NdFeAsO, although no definitive conclusion about the resistivity upturn can be provided at the moment. Close to room temperature, $NdFe_{1-x}Rh_xAsO$ compounds exhibit a Fermi-liquid behaviour, which denotes strong electron-electron interactions.

## Acknowledgements

This work was supported by the *Triangle de la Physique*, project STP 2008-095T. The authors acknowledge E. Rivière for SQUID measurements, F. Bouquet for his help with PPMS measurements and C. Godart for a gift of rhodium powder.